\documentclass[twocolumn,showpacs,prl,aps]{revtex4}
\usepackage{latexsym}
\usepackage{amssymb}
\usepackage{graphicx}
\usepackage{color}
\begin{document}

\bibliographystyle{prsty}

\title{Field Induced Staggered Magnetization and Magnetic Ordering in
Cu$_2$(C$_5$H$_{12}$N$_2$)$_2$Cl$_4$ }
\author{ M. Cl\'emancey$^1$, H. Mayaffre$^1$, C. Berthier$^{1,2}$,
M. Horvati\'c$^{2}$,\\ J.-B. Fouet$^3$, S. Miyahara$^4$, F.
Mila$^5$, B. Chiari$^6$, and O. Piovesana$^6$ }

\affiliation{$^1$Laboratoire de Spectrom\'etrie Physique,
Universit\'e J. Fourier \& UMR5588 CNRS, BP 87, 38402, Saint Martin
d'H\`{e}res, France}

\affiliation{$^{2}$Grenoble High Magnetic Field Laboratory, CNRS, BP
166, F-38042 Grenoble Cedex 09, France} \affiliation{$^{3}$Institut
Romand de Recherche Num\'erique en Physique des Mat\'eriaux,
PPH-Ecublens, CH-1015 Lausanne, Switzerland}
\affiliation{$^4$Department of Physics and Mathematics, Aoyama
Gakuin University, Sagamihara, 229-8558, Japan}
\affiliation{$^{5}$Institute of Theoretical Physics, Ecole
Polytechnique F\'ed\'erale de Lausanne, CH-1015 Lausanne,
Switzerland} \affiliation{$^{6}$Dipartimento di Chimica,
Universit\'a di Perugia, I-06100 Perugia, Italy}
\date{\today}

\begin{abstract}
We present a $^2$D NMR investigation of the gapped spin-1/2
compound Cu$_2$(C$_5$H$_{10}$N$_2$D$_2$)$_2$Cl$_4$. Our
measurements reveal the presence of a magnetic field induced
transverse staggered magnetization (TSM) which persists well below
and above the field-induced 3D long-range magnetically ordered
(FIMO) phase. The symmetry of this TSM is different from that of
the TSM induced by the order parameter of the FIMO phase. Its
origin, field dependence and symmetry can be explained by an
intra-dimer Dzyaloshinskii-Moriya interaction, as shown by DMRG
calculations on a spin-1/2 ladder. This leads us to predict that
the transition into the FIMO phase is not in the BEC universality
class.
\end{abstract}

 \maketitle

%%%%%%%%%%%%%%%%%%%%%%%%%%%%%%%%%%%%%%%%%%%%%%%%%%%%%%%%%%%%%%%%%%%%%%%%%%%%%%%%%%%%%%%%%%

Since the pioneering work of Haldane for S=1 antiferromagnetic (AF)
chains \cite {Haldane_1983}, the existence of collective singlet
ground states
%(valence bond crystals),
separated by an energy gap from the first triplet excited states
has triggered a very large number of theoretical and experimental
studies in low-dimensional quantum AF systems, as well as in 3D AF
coupled dimers \cite{QSS04}. Applying an external magnetic field
$H$ lowers the energy of the $M_S$ = -1 component of the triplet
band, inducing at some critical field value $H_{c1}$ a quantum
phase transition from a non-magnetic phase to a field-induced 3D
long-range magnetic ordered (FIMO) ground state, which has been
recently described in several cases as a Bose-Einstein
condensation (BEC) of triplet excitations
\cite{Affleck_1991,giamtsvelik_1999,Nikuni_2000,Jaime_2004}.
However, anisotropic terms in the spin-Hamiltonian, like
Dzyaloshinskii-Moriya (DM) interactions or staggered $g$ tensors,
often open a gap at $H_{c1}$, and change the universality class of
the transition \cite{Affleck_1991,Glazkov_2004,Sirker_2004}. Among
numerous
spin-liquid systems, %the organo-metallic compound
Cu$_{2}$(C$_{5}$H$_{12}$N$_{2}$)$_{2}$Cl$_{4}$ (Cu(Hp)Cl in short)
\cite{Chiari_1990} has long been considered as the prototype of a
$S = 1/2$ two-leg spin ladder in the strong coupling limit
$(J_{\perp} \gg J_{\parallel})$ \cite{ChaboussantEPJ_1998}.  Its
phase diagram in the $H$--$T$ plane consists of a gapped spin
liquid phase below $H_{c1}$, a 3D ordered magnetic phase between
$H_{c1}$~= 7.5~T and $H_{c2}$~= 13~T, and a  fully polarized,
gapped phase above $H_{c2}$. In the intermediate phase, between
$H_{c1}$ and $H_{c2}$, the interactions between the "magnetic
ladders" lead to the FIMO phase
\cite{Hammar,Calemczuck,Mayaffre_2000}. Close to the first quantum
critical point ($H=H_{c1},~T = 0$), $T_{\mathrm{FIMO}}(H)$ varies
as $(H-H_c)^{1/\alpha}$, where $\alpha$ was found to be close to
3/2, as expected in a BEC description \cite{giamtsvelik_1999}.
However, inelastic neutron scattering measurements
\cite{Stone_2002} have shown that the exchange paths are not
clearly identified in this system and are more complicated than in
the ideal picture described above.

%\section {Experiments}
 In this paper, we present a $^2$D NMR study of
a Cu(Hp)Cl single crystal (0.35$\times$0.3$\times$0.15 mm$^3$ size),
in which the N-H groups have been replaced by N-D groups. $^2$D
spectra have been recorded in the field range 3.5-15~T, and in the
temperature range 50$~$mK-50$~$K. From the study of the quadrupolar
couplings as a function of $T$ and $H$, we rule out previous
interpretations of the origin of FIMO as being magnetoelastic
\cite{Calemczuck,Mayaffre_2000}. We demonstrate the existence at low
$T$ of a sizable field-induced transverse staggered magnetization
(TSM) which starts to grow several tesla below $H_{c1}$, remains
nearly constant within the FIMO phase, and decays within a few tesla
above $H_{c2}$. Within the FIMO, it coexists with another TSM of
different symmetry corresponding to the order parameter of the FIMO.
Such a behavior has never been reported so far. We show that the TSM
outside the FIMO is well explained by Density Matrix Renormalization
Group (DMRG) calculations on a $S = 1/2$ spin ladder in the strong
coupling limit if
 a DM interaction is introduced on the rungs.

Selective deuteration provides several advantages over  proton
NMR. First, it divides by 6 the number of sites contributing to
the spectra. There are only 8 inequivalent sites for an arbitrary
orientation of $H$, and 4 for $H$ $\|$ $\vec{b}$ or in the
($\vec{a}, \vec{c}$) plane, due to the symmetry of the space group
P2$_1$/c. Secondly, the substituted $^1$H are the closest to Cu
atoms. Half of them are involved in the presumed exchange path
$J_{\parallel}$. These $^1$H sites show an anomalous shift at low
$T$, which has been previously considered as a hint for the
magnetoelastic character  of the transition into the FIMO
\cite{Mayaffre_2000}. Finally, $^2$D has a spin 1 and a
quadrupolar moment. Therefore, each inequivalent $^2$D site gives
rise to two lines. Their average position depends on the magnetic
hyperfine shift $K$ due to the coupling with the electronic spins
borne by the Cu$^{2+}$ atoms through the hyperfine tensors. Their
splitting $\delta\nu_Q$, due to the quadrupolar coupling with the
environment, is very sensitive to atomic displacements or
structural distortions. Both $K$ and $\delta\nu_Q$ strongly depend
on the orientation of $H$ with respect to the crystalline axes.
So, from the simultaneous determination of $\delta\nu_Q$ and $K$
versus $H$ and $T$, one can decide whether the variations of $K$
are due to atomic displacements or not. As a drawback, $^2$D NMR
is much less sensitive than $^1$H NMR. All spectra were obtained
by sweeping the frequency at fixed $H$ value and summing the
Fourier transforms of the echo \cite{Clark}. As long as the
hyperfine and the quadrupolar couplings are small perturbations
with respect to the Zeeman energy,
 \begin{eqnarray}
 \delta \nu_{\pm}^i =~^{2}\!\gamma \delta h^i_z \pm \delta\nu_Q^i/2
 \end{eqnarray}
where $\delta h^i_z = K^i H = \sum_{l,\alpha,\beta}
A_{z\alpha}^{i,l}(g^{-1})^{l}_{\alpha\beta} \chi_{\beta z}^{l} H$.
$K^i$ is the hyperfine shift of the deuteron site $i$,
$\mathbf{A}^{i,l}$ the hyperfine field tensor between copper site
$l$ and deuteron $i$, while $\mathbf{g}^l$ and $\mathbf{\chi}^l$
are respectively the $g$ and the susceptibility tensors of copper
$l$. Both $\mathbf{A}^{i,l}$ and $\mathbf{g}^{l}$ are $T$
independent in the absence of structural change.

Before considering the $T$ and $H$ dependence of
$\mathbf{\chi}$(T), we first demonstrate that the quadrupolar
couplings are $T$ independent below 10 K and $H$ independent at $T
\sim 0$ (50~mK). Fig.~1 shows $^2$D spectra as a function of $T$
for $H =$ 7.75~T $\|$ [11$\overline{1}$]. In this particular
orientation, the
\begin{figure}
\includegraphics[width=1\linewidth]{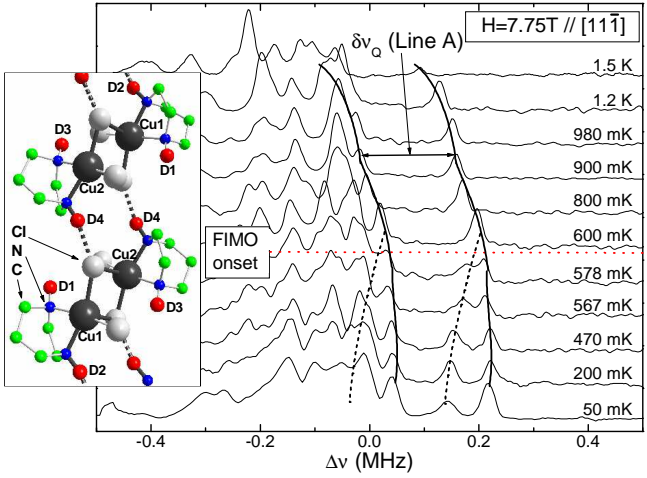}
\caption{(color online) $T$ dependence of $^2$D spectra for $H$ =
7.75~T $\parallel$ [11$\overline{1}$]. The quadrupolar doublet
corresponding to the line A at the right hand side of the spectra
is identified below 1.2 K. Its position is emphasized by solid
lines, which are just a guide for the eyes. Below 600 mK, each
component of the doublet splits into two lines (dashed versus
solid line), due to the entrance into the FIMO phase. Inset:
Structure of Cu$_2$(C$_5$H$_{12}$N$_2$)$_2$Cl$_4$. The 10 protons
attached to the C atoms are not shown. The other ladders can be
deduced by translation and rotation around the $b$ axis.}
\end{figure}
line A at the right hand side of the spectra remains isolated in
the whole $T$ range 50~K-50~mK. Several approaches have been used
to determine  the $T$ dependence of $\delta\nu_Q$ for this site.
Between 50 to 10 K, where all the $^1$H lines have a similar $T$
dependence, we made use of equation (1) by plotting $\delta
\nu_A(T)$ versus $^1$H shift $\delta \nu_H(T)$, which reflects
directly the variation of $\chi_S(T)$. The extrapolation to
$\delta \nu_H$ = 0 gives $\delta\nu_Q^A$~=~170~$\pm$~8~kHz. Below
1~K, the second line of the doublet involving line A also becomes
isolated in the spectra and the value of $\delta\nu_Q$ can then be
directly determined. It is found  constant through the 3D
transition and its  value 175 $\pm$ 6 kHz (Fig.~1 and 2a) is
consistent with that determined  by the high $T$ procedure. These
results strongly support the absence of structural transition in
the whole temperature range 50~K-50~mK, and in particular at
$T_{\mathrm{FIMO}}$.
\begin{figure}
\includegraphics[width=1\linewidth]{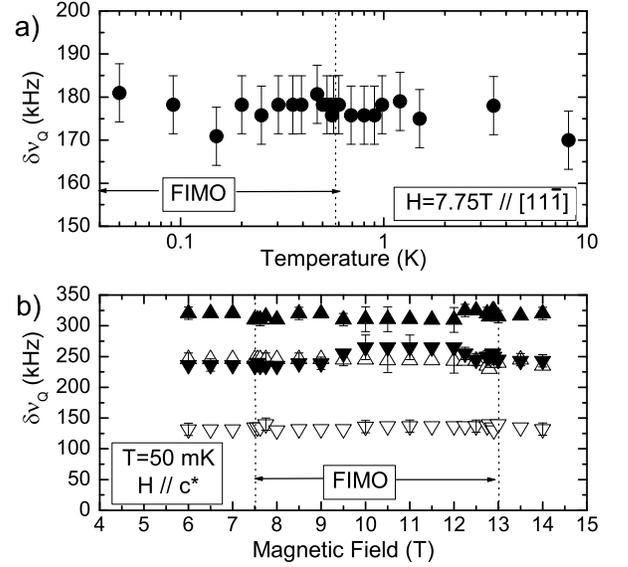}
\caption{a) $T$ dependence of $\delta\nu_Q$ for the site
corresponding to the line A for $H$ = 7.75~T and $\parallel$ to
[11$\overline{1}$] b) $H$ dependence of $\delta\nu_Q$ of the 4 sites
at 50~mK for $H~\|~c^*$. In both cases, $\delta\nu_Q$ remains
constant. These data demonstrate the absence of any structural
transition at the quantum critical fields $H_{c1}$ and $H_{c2}$ at
$T \sim 0$.}
\end{figure}
Let us now consider the $H$ dependence of $\delta\nu_Q$ at low
$T$. For that purpose, we used the spectra recorded at 50~mK in
the field range 6-14~T,  with $H$ $\|$ $c^*$. In this case the
symmetry reduces  the number of inequivalent $^{2}$D sites to 4,
and it is possible to fit the whole spectra. Fig.~2b shows the
values of $\delta\nu_Q$ for the 4 sites as a function of $H$. They
remain constant in the whole $H$ range, in particular throughout
the two quantum transitions at $H_{c1}$=7.5~T and $H_{c2} \simeq$
13~T. This rules out the existence of a structural transition at
$T$ = 50~mK associated with the quantum magnetic phase transition.
This  corrects our previous interpretation on the nature of the
FIMO \cite{Mayaffre_2000}, and contradicts the recently published
X-ray experiments \cite{Lorenzo}. Our experiments also show the
absence of any hysteresis upon entering the FIMO  by varying $H$
at fixed $T$ = 50~mK.
\begin{figure}
\includegraphics[width=1\linewidth]{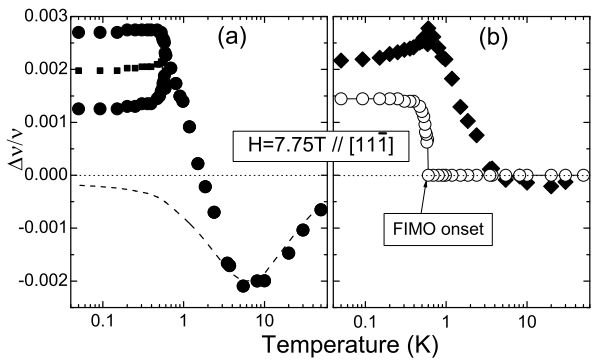}
\caption{a) Comparison between the $T$-dependence of
 $A_{zz}\chi_{zz}$ (dashed line) \cite{note_chi} and the
shift $K$ ($\bullet$) of the line A at $H$ = 7.75~T $\|$
[11$\overline{1}$], $A_{zz}$ = -470~G/$\mu_{\mathrm{B}}$. Below
$T_{\mathrm{FIMO}}$ = 600~mK, the line splits into two components,
and solid squares correspond to their average position. b) The
difference $A_{zz}\chi_{zz}(T) - K(T)$ ($\blacklozenge$) is due to
the onset of a transverse magnetization (see the text). The
splitting $S(T)$ ($\circ$) of the A line below $T_{\mathrm{FIMO}}$
is proportional to the order parameter of the FIMO. }
\label{figure3}
\end{figure}

We can now focus on the $T$ and $H$ dependence of the shift $K$
which directly reflects that of $\mathbf{\chi}$. In Fig. 3a, we
have  reported the $T$ dependence of the hyperfine shift $K
=\delta h_z /H$ of the  A line and of the product of the
longitudinal spin susceptibility $\chi_{zz}(T)$ by the hyperfine
field $A_{zz}$ in the orientation [11$\overline{1}$]
\cite{note_chi}. These two quantities are no longer proportional
below 3 K, and this can only be explained by the growing of a
transverse magnetization:
  \begin{eqnarray}
\delta h_{z}(T)= A_{zz}M_z(T) + \delta H_{tr}(T),
  \end{eqnarray}
where $\delta H_{tr} = A_{zx}M_x(T) + A_{zy}M_y(T)$. In Fig. 3b,
the $T$ dependence of the contribution of this transverse
magnetization to $K(T)$ is plotted. Below $T_{\mathrm{FIMO}}$ we
consider the mean position of the split line (solid squares) and
compare it to that of the splitting $S$ which is proportional to
the order parameter of the FIMO.

More information on $\delta H_{tr}$ can be obtained by looking at
its $H$ dependence  between 6 and 14~T at 50~mK, in the
orientation $H \parallel c^*$. The experimental values of the
shifts as a function of $H$ are shown in Fig. 4  and compared with
their expected variations in the absence of $\delta H_{tr}$, {\it
i.e.} $A_{zz}^i M_z(H)$. The solid and the dotted lines correspond
to the extremal values of $A_{zz}^i$ of the 4 hyperfine fields
determined between 50 and 10 K.
\begin{figure}[t]
\includegraphics[width=1\linewidth]{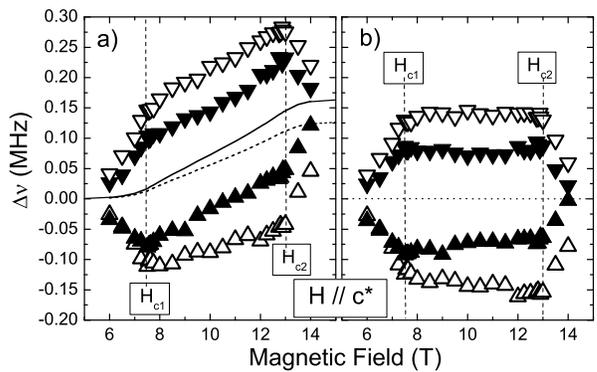}
\caption{ a) $H$ dependence of the hyperfine experimental shift of
the 4 sites ($\blacktriangledown, \blacktriangle, \triangledown,
\vartriangle$) for $H$ $\|$ $c^*$. Between $H_{c1}$ and $H_{c2}$,
the mean position of the split lines is plotted.  The solid and the
dotted lines are defined in the text. b) $H$ dependence of the
fraction of the shift due to the field induced TSM.}\label{fig_4}
\end{figure}
One clearly sees before $H_{c1}$ the growing of two by two opposite
values of $\delta H_{tr}$, which remain nearly constant within the
FIMO, and then start to decrease to zero. This immediately points to
the existence of a staggered magnetization, as in
SrCu$_2$(BO$_3$)$_2$ \cite{Kodama_2005}. However, it must be noticed
that for both orientations of $H$ with respect to the crystal, the
field induced TSM does not split the NMR lines, as clearly seen
outside the FIMO. This means that the number of inequivalent sites
remains the same outside  the FIMO (4 or 8 when $H$ $\|$  $c^*$ or
[11$\overline{1}$]). This differs from the effect of the spontaneous
TSM arising in the FIMO, which clearly splits them. The absence of
splitting due to the field induced TSM implies that this TSM
respects the full symmetry of the crystal space group, while that
corresponding to the order parameter of the FIMO does not, as
expected in a phase transition.

The progressive appearance of a TSM upon approaching  $H_{c1}$ at
low $T$ must come from interactions that break the SU(2) symmetry
and mix the $S = 0$ ground state with the lowest triplet $S = 1$
excitations. This can be due to DM interactions and/or staggered
$g$-tensors on dimers without inversion symmetry
\cite{Kodama_2005}, which is the case for the shortest
Cu$_1$-Cu$_2$ pairs \cite{Chiari_1990}. To model the effects of DM
interactions, we have performed DMRG calculations on a spin-1/2
ladder assuming a staggered distribution of DM vectors
$\mathcal{D}$ on the rungs, as the simplest model consistent with
the presence of an inversion center between neighboring dimers
\cite{footnote1}. Since the exchange paths are still debated,
other geometries including \textit{e.g.} diagonal inter-rung
couplings could in principle be realized. However, preliminary
results show that, as far as the TSM perpendicular to
$\mathcal{D}$ is concerned, the field dependence is essentially
the same up to an overall factor \cite{Fouet_unpublished}. As seen
in Fig.~5a, the results reproduce the experimentally observed TSM
fairly well for reasonable ratios $\mathcal{D}$/$J_\perp$ =0.05
and $J_{\parallel}/J_\perp$=0.2, with $J_\perp$ = 13~K.
Furthermore, since $\vec{m_{\bot}}$ $\propto$ $\vec{H}\times \vec{
\mathcal{D}}$ \cite{Fouet_unpublished}, it is easy to show,
considering the symmetry transformations of a pseudo-vector in the
group P2$_1$/c, that the field induced TSM due to the DM
interaction cannot split the $^2$D lines, whatever  the
orientation of $H$: the alternation of the DM vector from one rung
to the adjacent one corresponds to the symmetry of the lattice so
that, if $\vec{S_1}$ and $\vec{S_2}$ correspond respectively to
Cu$_1$ and Cu$_2$ electronic spins, $\vec{\mathcal{D}} \cdot
\vec{S_1} \times \vec{S_2}$ remains invariant.

\begin{figure}[t]
\includegraphics[width=1\linewidth]{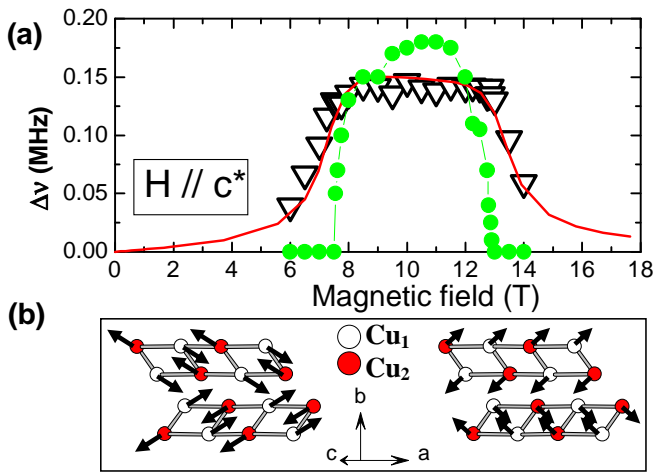}
\caption{(color online)  a) $H$ dependence of the field induced
TSM ($\triangledown$) and the fit (solid line) by the DMRG
calculation with $J _\perp= 13$~K, $J_\parallel/J_\perp=$~0.2 and
$\mathcal{D}/J_\perp =$~0.05. The order parameter of the FIMO is
proportional to $S(H)$ (solid circles). b) Schematic
representation of the two "ladders" and  the Cu(1) and Cu(2) sites
viewed from the $c^{\star}$ direction. Note an inversion symmetry
between two successive rungs. Left: Schematic representation of
the field induced TSM. Its direction on one rung is fixed by
$\vec{H}\times \vec{ \mathcal{D}}$ and the whole pattern can be
deduced by applying the symmetry operations of the lattice. Right:
TSM pattern consistent with the order parameter of the FIMO. The
direction in the plane is unknown and fixed by the inter-dimer
interactions.} \label{fig_5}
\end{figure}

The situation is different for the TSM corresponding to the order
parameter of the FIMO, which splits all the $^2$D lines whatever
the field orientation. This means that it breaks the inversion
symmetry between two adjacent rungs. An example of a possible
pattern is shown in Fig.~5b. This would be consistent with a BEC
in weakly coupled dimers systems into a uniform phase in which the
TSM is the same on all rungs. However, the BEC description is here
inadequate since, due to the field induced TSM, the isotropy in
the plane perpendicular to $H$ is lost. So, the ordered phase can
only break a discrete symmetry, and the phase transition into the
FIMO cannot fall into the universality class of BEC. This is
actually supported by the present data. Indeed, $T_c(H)$ varies as
($H-H_{c1}$)$^{0.55}$, consistent with the results reported in
\cite{Mayaffre_2000}, but in slight disagreement with the BEC
prediction ($H-H_{c1}$)$^{2/3}$ [\onlinecite{footnote}]. Moreover,
the field dependence of the order parameter within the FIMO, as
revealed by the splitting $S$, was found to be
($H-H_{c1}$)$^{\beta}$ with $\beta$~=~0.34~$\pm$~0.04
\cite{these_martin}, instead of the mean-field value $\beta~=~0.5$
expected for a BEC in 3+1 dimensions. Unfortunately, the exponent
for the transition as a function of $T$ at fixed $H$ could not be
reliably extracted from the present data. So, the precise
determination of the nature of the phase transition into the FIMO
is left for future investigation. Finally, the presence of DM
interactions is normally accompanied by a gap opening,
corresponding to some kind of level anticrossing
\cite{Kodama_2005}. In Cu(Hp)Cl, a fast decrease of the nuclear
spin-lattice relaxation $T_1^{-1}$ has indeed been observed below
1~K in the field range of the FIMO, even at $H$ = 7.75~T for which
$T_{\mathrm{FIMO}}$ = 300~mK \cite{Mayaffre_2000}. These data are
consistent with the gap estimated by DMRG
\cite{Fouet_unpublished}.

In conclusion, we have shown that in Cu(Hp)Cl two types of
transverse staggered magnetization appear. The first one, which
extends outside the 3D FIMO phase, is field induced and breaks the
SU(2) symmetry, but not the crystal symmetry. We propose that it
is due to the presence of a DM interaction $\mathcal{D}$ on the
strong dimers, and we have shown by an explicit calculation in a
ladder geometry that a quantitative fit can be achieved for a
reasonable value of the parameters
($\mathcal{D}/J_\perp~\simeq~0.05$). The second one, which
corresponds to the order parameter of the FIMO, breaks the
inversion symmetry of the structure. We think that these features
are generic for an assembly of interacting non centro-symmetric
dimers, and we suggest that they will change the universality
class of the field induced phase transition.

\begin{acknowledgments}
We acknowledge useful discussions with A. L\"auchli and with K.
Penc. This work was supported by the Swiss National Fund and by
MaNEP.
\end{acknowledgments}

%%%%%%%%%%%%%%%%%%%%%%%%%%%%%%%%%%%%%%%%%%%%%%%%%%%%%%%%%%%%%%%%%%%%%%%%%%%%%%%%%%%%%


\begin{thebibliography}{10}
\bibitem{note_chi} For this field value and orientation, $\chi_{zz}(T)$ has been extrapolated  from the
 data in Ref \cite{chaboussantthesis}. Below 1.4 K, $\chi_S$ is smoothly
extrapolated to its value corresponding to the positive value of
$M/H$ at $T$ = 0.
\end{thebibliography}
\end{document}